\documentclass[aps,pra,reprint,superscriptaddress]{revtex4-1}
\usepackage{amsmath}
\usepackage{graphicx}
\usepackage{color,longtable}
\usepackage{epstopdf}

\newcommand{\ket}[1]{| {#1} \rangle}
\newcommand{\expect}[1]{\langle {#1} \rangle}

\begin{document}

\title {Quantum Interferometry with Microwave-dressed F=1 Spinor Bose-Einstein Condensates: Role of Initial States and Long Time Evolution}

\author{Qimin Zhang}
\email{qmzhang@ou.edu}
\author{Arne Schwettmann}
\email{schwettmann@ou.edu}
\affiliation{Homer L. Dodge Department of Physics and Astronomy, The University of Oklahoma, Norman, USA}

\begin{abstract}
We numerically investigate atomic interferometry based on spin-exchange collisions in $F=1$ spinor Bose-Einstein condensates in the regime of long evolution times $t\gg h/c$, where c is the spin-dependent interaction energy. We show that the sensitivity of spin-mixing interferometry can be enhanced by using classically seeded initial states with a small population prepared in the $m_F=\pm1$ states.
\end{abstract} 

\maketitle

\section{Introduction}

In a spinor Bose-Einstein condensate (BEC), the atomic hyperfine spin degree of freedom becomes accessible and displays fascinating quantum dynamics driven by collisions that can be controlled via external fields. Spin-exchange collisions in $F=1$ microwave-dressed spinor BECs, where two atoms in $m_F=0$ Zeeman substates collide with each other and change into a pair of entangled atoms in $m_F=\pm1$ states, create a rich dynamical system with analogies to four-wave mixing in atomic vapors~\cite{Marino2018}, the bosonic Josephson effect~\cite{Stenholm1999}, the quantum non-rigid pendulum~\cite{Law1998}, and with quantum phase transitions that can lead to creation of massive entanglement~\cite{You2017}. The spin-exchange collisions conserve total spin and magnetization~\cite{Goldstein1999}. The collisions cause characteristic population oscillations between the $m_F=0$ and the $m_F=\pm1$ states~\cite{Romano2004,Pu1999,Chang2005} and can generate squeezing~\cite{Duan2002,Pu2000,Vogels2002}. Surprising phenomena that have been observed in spinor BECs driven by spin-exchange include spin textures and spin waves in elongated spinor BECs~\cite{Hansen2013,Gu2004}, spin dynamics in lattices~\cite{Widera2005,Zhao2015} and spin-nematic squeezing~\cite{Hamley2012a}.

It was demonstrated that spin dynamics can be precisely controlled using microwave dressing~\cite{Zhao2014} and, recently, a phase-sensitive amplifier was implemented using this control~\cite{Wrubel2018}. This opens up the field of matter-wave quantum optics in spin space. In particular, quantum interferometry with sensitivities beyond the standard quantum limit (SQL), based on spin-exchange collisions, is possible. So far, experiments on quantum interferometry in this system started with all atoms in $m_F=0$ and allowed only a few atoms to populate the arms of the interferometer during the evolution~\cite{Linnemann2016}. Here, we are interested in quantum interferometry starting with initial states where some atoms are seeded in $m_F=\pm1$. In addition, we investigate the effect of long evolution times with more than a few atoms in the arms of the interferometer, beyond the regimes of validity of the Bogoliubov, truncated Wigner, and undepleted pump approximations.

The investigations in this paper focus on numerical simulations of the collisional evolution of spin populations in a $F=1$ sodium BEC. We simulate a nonlinear spin-exchange based interferometer that measures the relative phase between $m_F=0$ and $m_F=\pm1$ pairs. The phase measurement exhibits uncertainties that improve upon the SQL. We focus on quantum-enhanced interferometry where there are macroscopic numbers of atoms in the arms of the interferometer. This is desirable compared to small populations, because it makes detection easier in experiments. This regime can be realized via long evolution times where many collisions are allowed to take place, and via populating the $m_F=\pm1$ states initially, which can speed up the evolution. We show that there are parameter regimes in which such an interferometer can surpass the SQL. The interferometer fringes become highly non-sinusoidal, owing to the nonlinear nature of the phase measurement.

\section{Computational method}

We consider small F=1 BECs where the Thomas-Fermi radius is smaller than the spin healing length, ${\xi_s=2\pi\hbar/\sqrt{2m|c_2|\overline n}}$ and spin domain formation is therefore energetically suppressed. Here, ${c_2=4\pi\hbar^2(a_2-a_0)/3m}$, with $a_0$ and $a_2$ the scattering lengths for the two allowed collision channels of total spin 0 and 2 \cite{Knoop2011}, $m$ is the atomic mass, and $\overline n$ is the mean number density~\cite{Zhang2005}. We assume further that the spin-dependent interaction is much weaker than the density-dependent interaction. This allows us to make the single-spatial-mode approximation (SMA), which assumes that all spin components share the same spatial wavefunction~\cite{Yi2002,Zhang2005}. Under the SMA, the evolution is governed only by the spin part of the Hamiltonian, $H_s$. In the presence of microwave-dressing and an applied magnetic field~\cite{Stenger1998},
\begin{equation}
\hat H_s=\frac{c}{2N}\hat{\bf F}^2-q\hat a_0^{\dagger} \hat a_0,
\end{equation}
where $\hat{\textbf{F}}=a_\alpha^\dagger\textbf{F}_{\alpha \beta}a_\beta$ is the total spin operator, and $\textbf{F}_{\alpha \beta}$ are spin-1 matrices. Here, $\mbox{$c=c_2\overline n$}$ is the spin-dependent interaction parameter. In a typical small sodium spinor BEC in a crossed far-off resonance trap with geometric mean trap frequency of $200$ Hz and ${N\approx75{,}000}$, we have $c/h\approx30$ Hz~\cite{Liu2009}. q is the effective quadratic Zeeman shift, ${q/h\approx \gamma B^2-\frac{\Omega^2}{\Delta_\mu}}$, where $\gamma B^2$ is the quadratic Zeeman shift due to the applied magnetic field $B$, and ${\gamma\approx277~\textrm{Hz/G}^{2}}$ for sodium~\cite{Stenger1998}, $\Omega$ is the microwave Rabi frequency on resonance, and $\Delta_{\mu}$ is the detuning from the $\ket{F=1, m_F=0}\rightarrow \ket{F=2, m_F=0}$ transition. Here, we assumed that ${\Delta_\mu\gg \gamma}$. q can be used to control the spin dynamics via the magnetic field or the microwave dressing.
We simulate the evolution according to $\hat H_s$ using two numerical methods: the full quantum evolution and the truncated Wigner approximation. These two methods are contrasted in the following sections.

\subsection{Full quantum evolution}
The full quantum method consists of calculating the time evolution propagator $e^{-i\hat H_st}$ of the system in the basis of Fock states  $\ket{N_{-1},N_0,N_{+1}}$, where $N_i$ is the occupation number of the i-th magnetic sublevel. We use the Chebyshev propagator to solve this quantum mechanical time evolution numerically on a supercomputer. Compared to other methods, such as the second-order difference (SOD) method~\cite{Kosloff1983} and the short-iterative Lanczos (SIL) method~\cite{Park1986}, the Chebyshev propagator is more accurate and efficient, and requires much less memory and CPU time~\cite{Chen1999}. For Hermitian Hamiltonians~\cite{Chen1999}, 
\begin{equation}
e^{-i\hat Ht}=\sum_{k=0}^{\infty} (2-\delta_{k0})(-i)^k J_k(t)T_k(\hat H),
\end{equation}
where $J_k$ are Bessel functions of the first kind, $T_k(\hat H)$ are Chebyshev polynomials, and $\hat H$ is the Hamiltonian scaled to [-1,1]. The Chebyshev propagator can be calculated recursively and precisely because it consists of polynomials of $\hat H$ that obey simple recursion relations, compared to evolution via the exponential function which is harder to compute directly. The recursion relations we use are~\cite{Chen1999}
\begin{equation}
T_{k+1}(\omega)=2\omega T_k(\omega)-T_{k-1}(\omega),\, \textnormal{for}\;  k\geq1,
\end{equation}
with
\begin{equation}
T_0(\omega)=1,  \; T_1(\omega)=\omega.
\end{equation}

This method can be used for arbitrary initial state, and we focus on two kinds of initial states: pure Fock states  $\ket{N_{-1},N_0,N_{+1}}$ with fixed number of atoms in each state and spin coherent states $\ket {\alpha_{-1}, \alpha_0, \alpha_{+1}}=\sum\limits_{N_{-1}, N_0, N_{+1}=0}^N \sqrt{\frac {N!}{N_{-1}!N_0!N_{+1}!}}\alpha_{-1}^{N_{-1}}\alpha_0^{N_0}\alpha_{+1}^{N_{+1}}\ket{N_{-1},N_0,N_{+1}}$, where $\alpha_i=\sqrt{\expect{N_i}}e^{i\expect {\theta_i}}$ with mean population $\expect{N_i}$ and phase $\expect{\theta_i}$. The magnetization ${M=N_{+1}-N_{-1}}$ is fixed in a Fock state but ranges from -$N$ to $+N$ in a spin coherent state. The total atom number ${N=N_{-1}+N_0+N_{+1}}$ is conserved in both cases, and constrains the sum for the coherent states. Due to conservation of total atom number ${N=N_{-1}+N_0+N_{+1}}$ and magnetization ${M=N_{+1}-N_{-1}}$, the Fock basis $\ket{N_{-1},N_0,N_{+1}}$ can also be expressed as $\ket{\frac{1}{2}(N-N_0-M),N_0,\frac{1}{2}(N-N_0+M)}$. The computation for a Fock initial state is much faster than that for a coherent initial state, because of the limited subspace of allowed occupation numbers.\par

\subsection{Truncated Wigner approximation (TWA)}
In some calculations, we use a semi-classical approach based on the truncated Wigner and mean-field approximations to approximate the full quantum spinor dynamics. In this method, the interactions between each atom and all other atoms during spin collisions are treated as an average interaction. The Hamiltonian is thus simplified as
\begin{equation}
\hat {H}_s^{\textrm{TWA}}=\hbar c(\expect{\hat F_x}\hat F_x+i\expect{\hat F_y}\hat F_y+\expect{\hat F_z}\hat F_z)+\hbar q\hat F_z^2,
\end{equation}
Here, $\hat F_{\alpha}$ are spin-1 matrices in the basis $\ket{F, m_F}$.
We set the initial state to approximate a three-mode coherent spin state with standard deviations of $\sigma_{N_i}=\sqrt{\frac{1}{4}+\expect{N_i}}$, as
\begin{equation}
\mathbf{\Psi}=\mathbf{\Psi_0}+\mathbf{\delta}\,\frac{1}{2\sqrt{N}},
\end{equation}
where
\begin{equation}
\mathbf{\Psi_0}=
\left(
\begin{array}{c}
\psi_{-1}\\
\psi_0\\
\psi_{+1}
\end{array}
\right)
=
\left(
\begin{array}{c}
\sqrt{\frac{\expect{N_{-1}}}{N}}\,e^{i\expect{\theta_{-1}}}\\
\sqrt{\frac{\expect{N_{0}}}{N}}\,e^{i\expect{\theta_0}}\\
\sqrt{\frac{\expect{N_{+1}}}{N}}\,e^{i\expect{\theta_{+1}}}\\
\end{array}
\right),
\end{equation}
and
\begin{equation}
\mathbf{\delta}=
\left(
\begin{array}{c}
a+i\,b\\
c+i\,d\\
f+i\,g
\end{array}
\right),
\end{equation}
where a, b, c, d, f and g are real random numbers, drawn independently from a normal distribution with zero mean and a standard deviation of 1.  $\expect{\theta_i}$ are the mean phases, $\expect{N_{+1}}$, $\expect{N_{-1}}$ are the initial mean seed populations, $N$ is total atom number, and $\expect{N_0} = N - \expect{N_{+1}} - \expect{N_{-1}}$ is the initial number of $m_F=0$ atoms (before addition of noise). We define the spinor phase $\theta=\theta_{+1}+\theta_{-1}-2\theta_{0}$. Setting the initial spinor phase in $\psi_0$ is accomplished by letting $\expect{\theta_{+1}} = \expect{\theta_{-1}} = 0$ so that $\expect{\theta_0}=-\expect{\theta}/2$. In all the simulations presented here, we set $\expect{\theta_0}=0$. \par

The evolution is then calculated by propagating the effective single particle wavefunction via ${\Psi(t+dt)=\exp{({{-i\hat{H}_s^\textrm{TWA}} dt})}\Psi(t)}$ and taking an ensemble average over many realizations. We found that the TWA works well for short and intermediate evolution times compared to $h/c$ when starting with all atoms in $m_F=0$. The TWA fails to predict correct standard deviations when starting with some seeded atoms in $m_F=\pm1$ and when the evolution times become longer, $t\gg h/c$.

\subsection{Interferometer}

\begin{figure}[h!]
\centering
\includegraphics{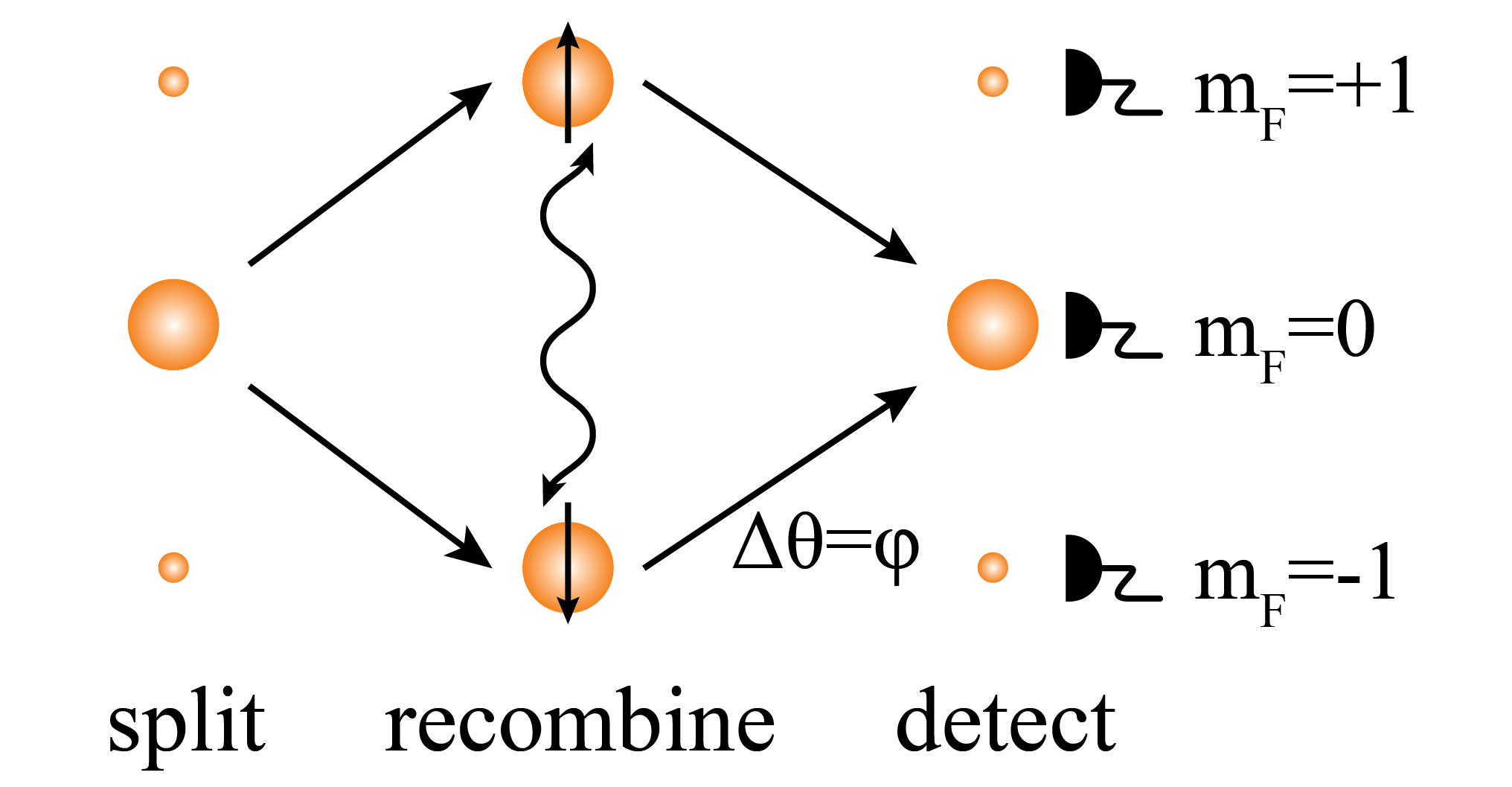}
\caption{(Color online) Cartoon of the interferometer sequence with initial seeds. The phase shift $\varphi=\Delta \theta$ is applied via microwave dressing. The straight arrows denote time evolution. The wavy arrow denotes entanglement. The black detectors represent population measurements via Stern-Gerlach time-of-flight absorption imaging at the end of the sequence.}
\label{fig:interferometry}
\end{figure}

We realize a spin-mixing interferometer sequence in three steps, similar to recent experiments and theoretical proposals~\cite{Linnemann2016, Gabbrielli2015}, as shown in Fig.~\ref{fig:interferometry}. Initially, we prepare $N$ atoms with certain classical seeds in $m_F=+1$ and/or  $m_F=-1$. In an experiment, the seeding can be done either via short resonant microwave pulses to transfer populations through an intermediate F=2 state, or via resonant rf pulses that transfer atoms directly from $m_F=0$ to $m_F=\pm1$. Following the initial state preparation, we let the system evolve for time $\tau$, after which there is a certain number of atoms ${\expect{N_{\textrm{inside}}}=\expect{N_{+1}}+\expect{N_{-1}}}$ in the $m_F=\pm1$ states. At time $\tau$, we apply a detuned microwave-dressing pulse with short duration, $t_{\textrm{rev}}\ll h/c$, and large amplitude, $q_{\textrm{rev}}\gg c$ and $q$. This pulse shifts the $m_F=0$ state and adds a phase shift $\varphi \approx2q_{\textrm{rev}}t_{\textrm{rev}}$ to the spinor phase $\theta$. We then let the system evolve for another time $\tau$, and  evaluate the final number of atoms in the $m_F=\pm1$ states, $N_+=N_{+1}+N_{-1}$, with mean value $\expect {N_+}$ and standard deviation $\sigma_{N_+}$ at time $t_f=\tau+t_{rev}+\tau$. In an experiment, detection can be done via Stern-Gerlach separation followed by time-of-flight absorption imaging. To characterize the phase sensitivity of such an interferometer, we analyze $\expect{N_+}$ and $\sigma_{N_+}$ as a function of $\varphi$ to find regions with the best sensitivity. The phase sensitivity is given by $(\Delta \varphi)^2=\frac{(\sigma_{N_+})^2}{|d\expect{N_+}/d \varphi|^2}$ from error propagation~\cite{Linnemann2016}. The SQL to be compared to $(\Delta \varphi)^2$ is defined as $\textrm{SQL}=1/\expect{N_{\textrm{inside}}}$ \cite{Linnemann2016}. Both sensitivity $(\Delta \varphi)^2$ and the SQL are determined by measuring the mean total population $\expect{N_+}$ in the $m_F=\pm1$ states and its standard deviation $\sigma_{N_+}$ at the end of the sequence. Our simulation codes were verified with known experimental results by reproducing Fig.~2(b) of Ref.~\cite {Linnemann2016} and Fig.~1 of Ref.~\cite{Pu1999}.

\begin{figure}[h!]
\centering
\includegraphics[width=0.9\linewidth]{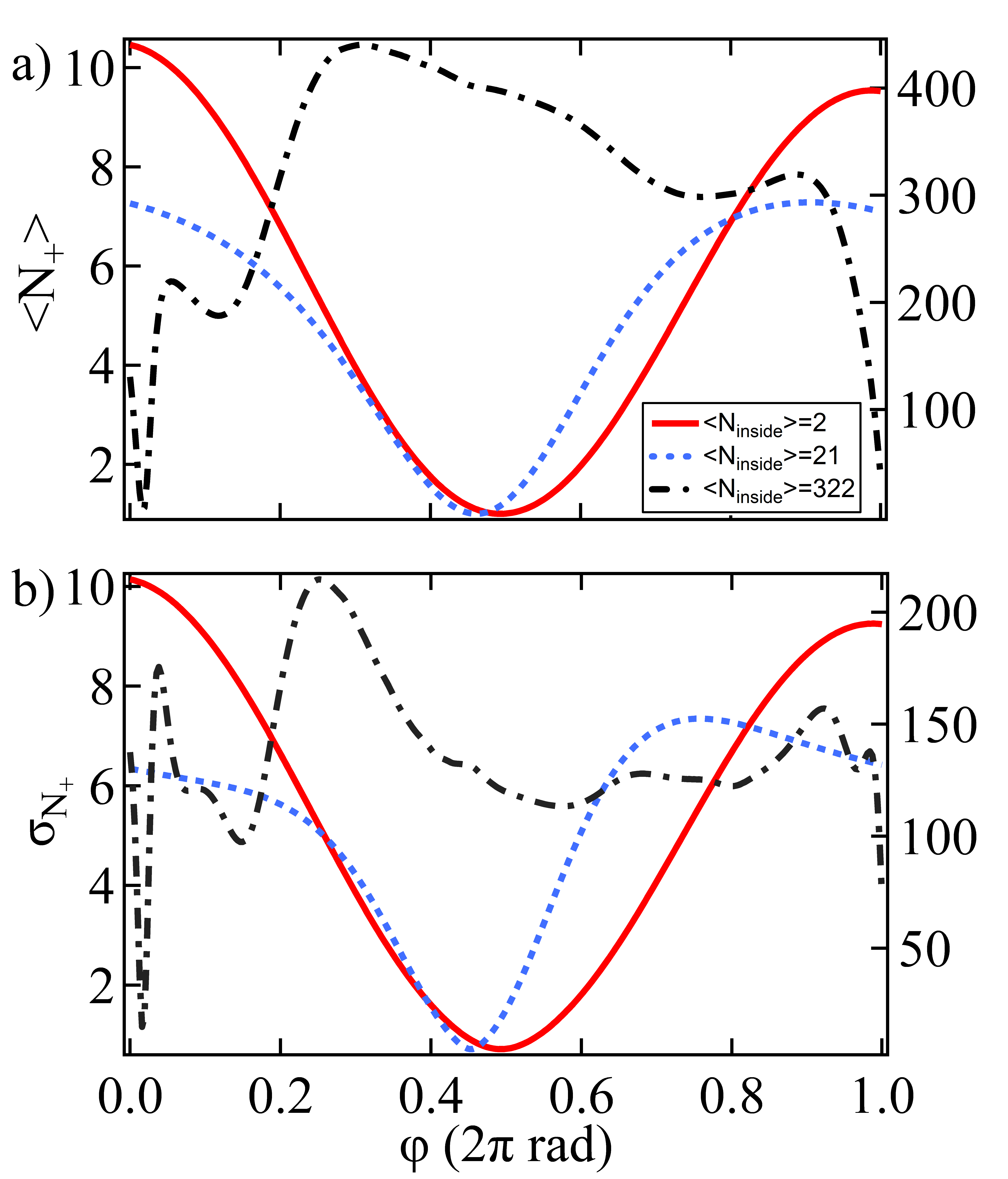}
\caption{(Color online) Interferometer fringes for different evolution times. Shown is the phase dependence of (a) $\expect{N_+}$ and (b) $\sigma_{N_+}$ for  $\expect{N_{\textrm{inside}}}=2$ (red solid, left axis), 21 (blue dashed, left axis), 322 (black dash-dotted, right axis). Here, ${c/h=30}$ Hz, ${q/h=-35}$ Hz, $N=1{,}000$, and zero initial seed. For longer evolution times (larger $\expect{N_{\textrm{inside}}}$), interferometer fringes become highly non-sinusoidal.} 
\label{fig:interferenceFringes}
\end{figure}

To investigate the role of the initial state and of long evolution times, we simulate the interferometry sequence starting from coherent initial states or Fock initial states with different initial seeds and $t_f\gg h/c$. We use realistic parameters for a sodium BEC~\cite{Knoop2011,Pechkis2013} with $c/h=30$ Hz, $q/h=-2$~Hz and $-35$~Hz, $q_{\textrm{rev}}/h$ ranging between 0~Hz and $-2{,}000$~Hz and $t_{\textrm{rev}}=0.25$~ms to achieve a phase shift of $\varphi=0\dotsc2\pi$. The initial spinor phase is set to $\expect{\theta}=0$. We choose different initial seeds to investigate the role of the initial state. An example of the effect of long evolution times is shown in Fig.~\ref{fig:interferenceFringes}. $\expect{N_+}$ and $\sigma_{N_+}$ vs.~phase $\varphi$ are sinusoidal only for short evolution times $\tau \ll h/c$ where $\expect{N_{\textrm{inside}}}\ll N$. At longer evolution times where $\expect{N_{\textrm{inside}}}$ is larger, $\expect{N_+}$ and $\sigma_{N_+}$ become highly non-sinusoidal. The non-sinusoidal depedence on phase can improve the interferometer sensitivity since $|d\expect{N_+}/d \varphi|$ can be enhanced.

\section{Results}
\subsection{Comparison of TWA evolution and Chebyshev evolution}

\begin{figure}[h!]
\centering
\includegraphics[width=0.9\linewidth]{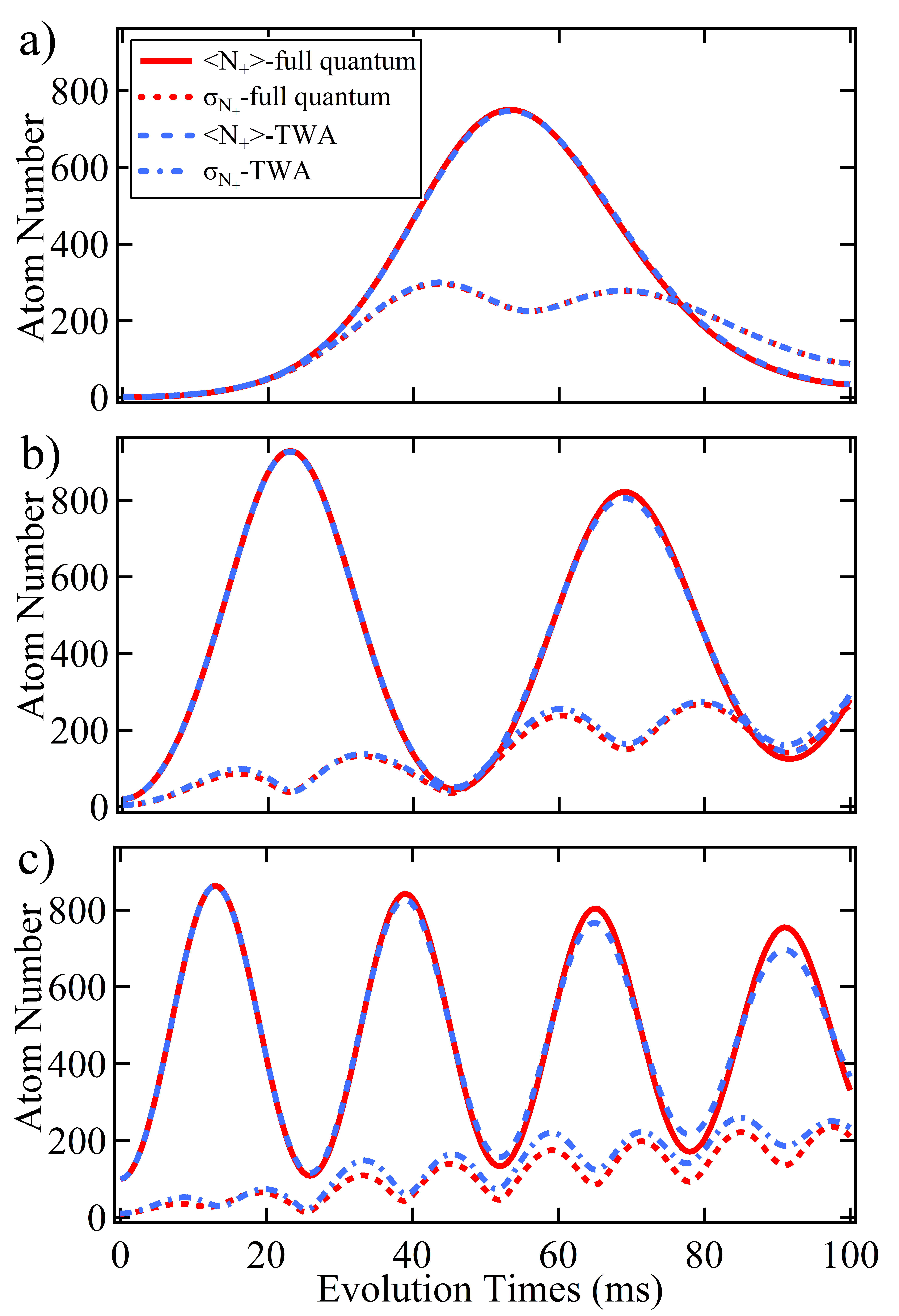}\label{fig:evolution_qeff2}
\caption{(Color online) Evolutions of $\expect{N_+}$ for the full quantum method (red solid) and TWA method (blue dashed), and evolutions of $\sigma_{N_+}$ for the full quantum method (red dotted) and TWA method (blue dash-dotted). Shown are evolutions for (a) 0\%, (b) 2\% , and (c) 10\% initial seeds. Here, $N=1{,}000$ and ${q/h=-2}$~Hz. For large initial seeds, the standard deviations predicted by the TWA method are in disagreement with the full quantum method.} 
\label{fig:evolution}
\end{figure}
To determine the range of validity of the TWA method, we compare the results from the TWA method with the full quantum method. We find that the standard deviations $\sigma_{N_+}$ predicted by the TWA method are only accurate for non-seeded evolutions. For seeded cases, they are only valid for short evolution times $t\ll h/c$, and then quickly diverge from the full quantum method. As shown in Fig.~\ref{fig:evolution}, the TWA method agrees well with the full quantum calculation for at least the first cycle of population oscillations in the unseeded case, see Fig.~\ref{fig:evolution}a. But as initial seeds are introduced into the system, the results from the TWA method no longer agree with the full quantum calculations, as seen in Fig.~\ref{fig:evolution}b and~\ref{fig:evolution}c. With initial seeds, the TWA method doesn't capture the quantum noise accurately anymore. Therefore, in this article, only the results for non-seeded evolutions were obtained using the TWA method, while all data for seeded evolutions were obtained using the full quantum method.

\subsection{Simulation for non-seeded initial states}

\begin{figure}[h!]
\centering
\includegraphics[width=\linewidth]{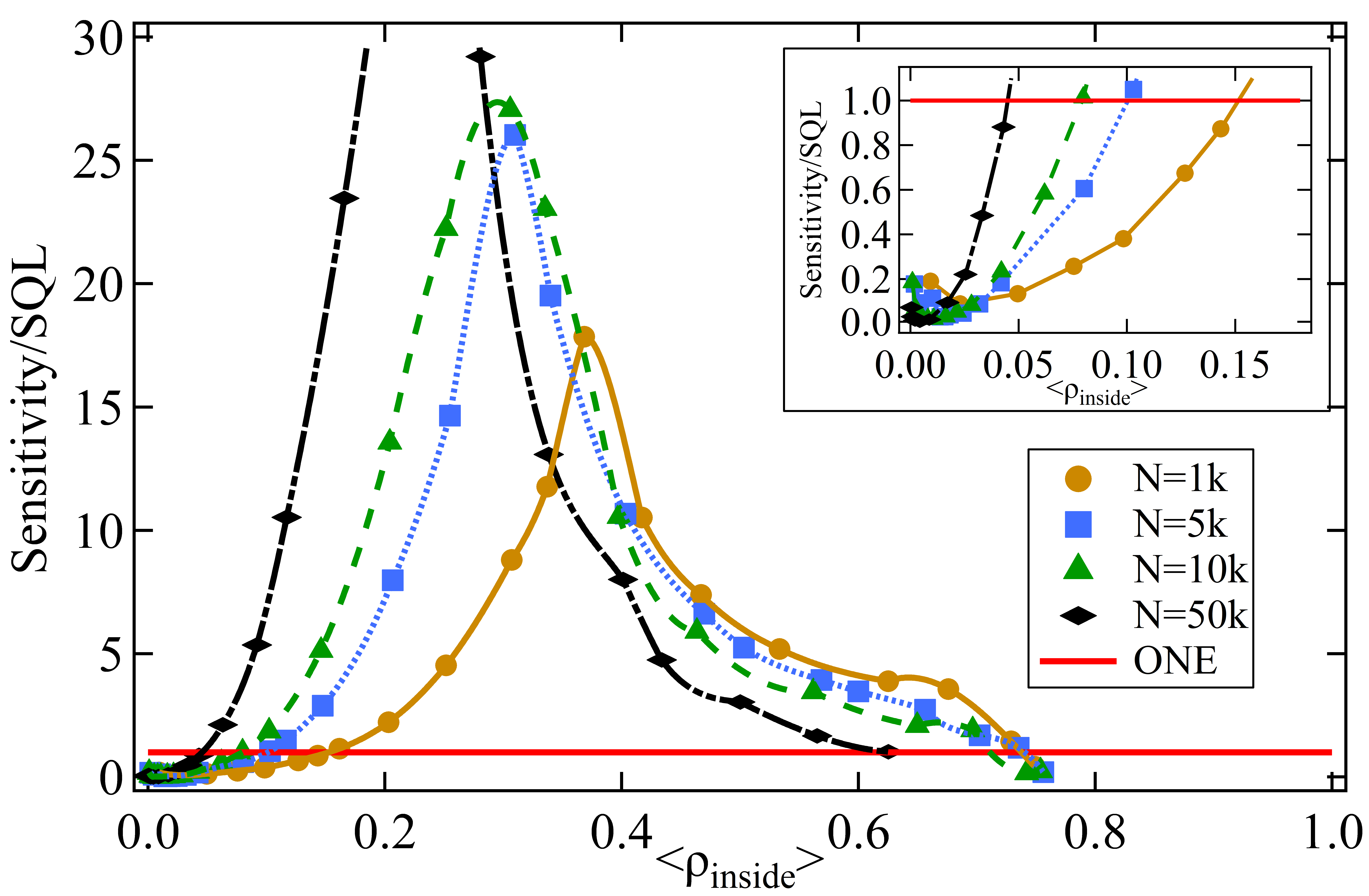}
\caption{(Color online) Phase sensitivities for different $N$ with zero initial seed. Shown are $N=1{,}000$ (yellow circles), $N=5{,}000$ (blue squares), $N=10{,}000$ (green triangles), $N=50{,}000$ (black diamonds). Here, $q/h=-2$ Hz. The red line depicts $\textrm{sensitivity/SQL = 1}$. Points below the red line correspond to quantum-enhanced sensitivity. The inset shows a zoomed-in region where enhanced sensitivities are found.  The lines are intended as guide to the eye.}
\label{fig:0seed}
\end{figure}

We first investigate the interferometer sensitivity and its dependence on the total number of atoms with a coherent initial state and zero initial seed. In order to find the best sensitivity for a given set of parameters, the best operating point of the interferometer is first determined. The best operating point is the phase shift $\varphi$ that minimizes $(\Delta \varphi)^2$. In Fig.~\ref{fig:0seed}, we plot the best sensitivities (lowest $(\Delta \varphi)^2$), normalized to the SQL, for different $N$ as a function of number fraction inside the arms of the interferometer $\mbox{$\expect{\rho_{\textrm{inside}}}=\frac{\expect{N_{\textrm{inside}}}}{N}$}$. From $N=1{,}000$ to $N=50{,}000$, the sensitivity/SQL ratio is similar and there are regions where the sensitivity beats the SQL (sensitivity/SQL $<$ 1) even for ${N=50{,}000}$. \par
To summarize, for a non-seeded spin-mixing interferometer, by going to long evolution times, we find sensitivities better than the SQL even with large total atom number $N=50{,}000$ and large numbers of atoms inside the arms of the interferometer $\expect{N_{\textrm{inside}}}>2{,}150$. 

\begin{figure}[h!]
\centering
\includegraphics[width=\linewidth]{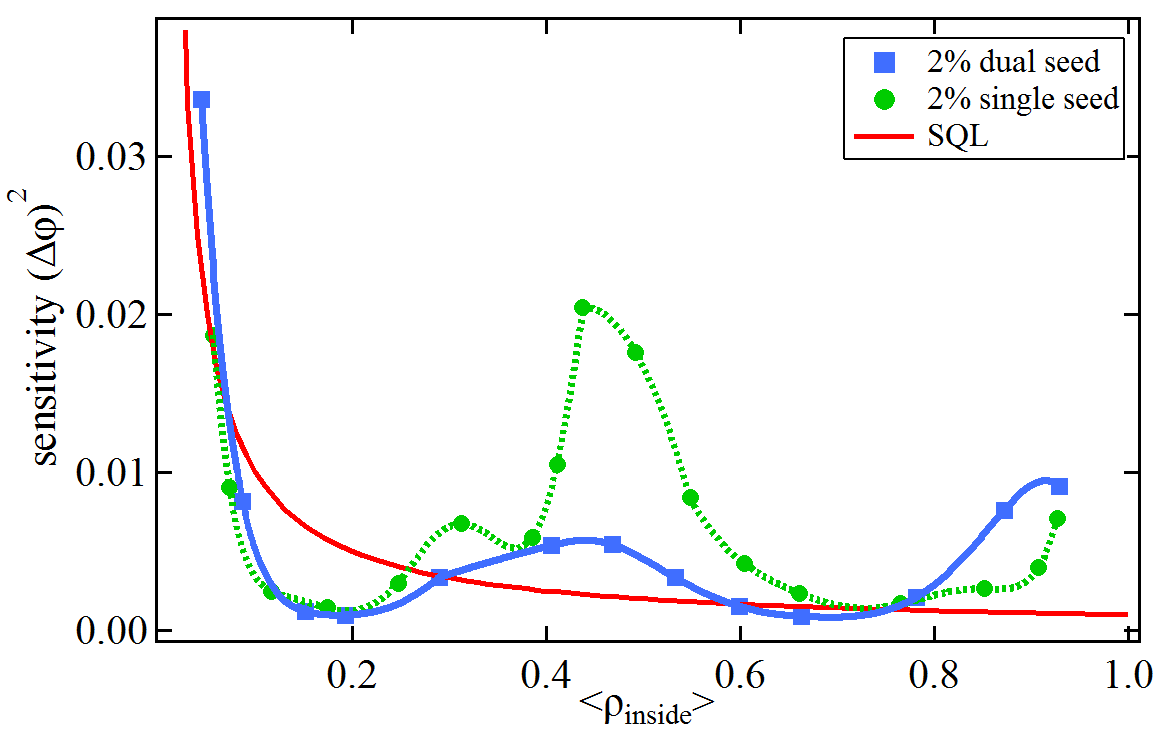}
\caption{(Color online) Phase sensitivities for dual initial seeds (blue squares) and single initial seeds (green circles) with $N=1{,}000$, coherent initial state and 2\% initial seeds. Here, $q/h=-2$~Hz. The red line depicts the SQL. Points below the red line correspond to quantum-enhanced sensitivities. The lines are intended as guide to the eye. }
\label{fig:dual-single}
\end{figure}

\subsection{Simulation for seeded initial states}

For evolutions with initial seeds, the initial seeds can be dual or single. For dual seeding, equal numbers of atoms are prepared in $m_F=+1$ and $m_F=-1$ states. For single seeding, all seeded atoms are prepared either in the $m_F=+1$ or in the $m_F=-1$ state. The effect of single and dual seeding on the phase sensitivity is shown in Fig.~\ref{fig:dual-single}. We observe quantum-enhancement for both types of seeds.

\begin{figure}[h!]
\centering
\includegraphics[width=\linewidth]{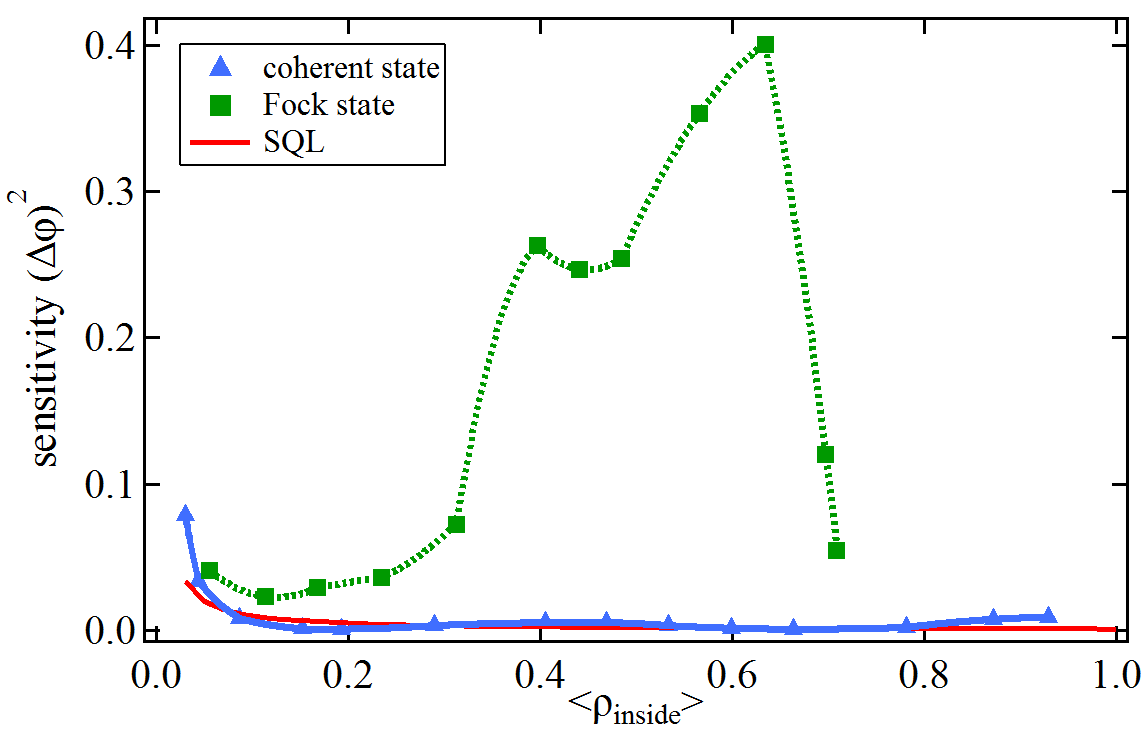}
\caption{(Color online) Phase sensitivities for coherent initial state (blue triangles) and Fock initial state (green squares) with ${N=1{,}000}$ and 2\% dual initial seeds. $q/h=-2$~Hz. The red line depicts the standard quantum limit. Points below the red line correspond to quantum-enhancement. The coherent initial state performs better than the Fock initial state for all values of $\expect{\rho_{\textrm{\textrm{inside}}}}$. The lines are intended as guide to the eye.}
\label{fig:coherent-Fock}
\end{figure}

The type of the initial state, either a coherent state or a Fock state, also makes a difference to the interferometer sensitivities. In Fig.~\ref{fig:coherent-Fock} we compare the sensitivities for a coherent initial state with those for a Fock initial state, for different $\expect{\rho_{\textrm{inside}}}$. Here, we set $N=1{,}000$ and used $2\%$ dual initial seeds. The interferometer with a coherent initial state has much better sensitivities than that with a Fock initial state. In the remainder of this article, all initial seeds are dual seeds, and all initial states are coherent states unless otherwise specified.

\begin{figure}[h!]
\centering
\includegraphics[width=\linewidth]{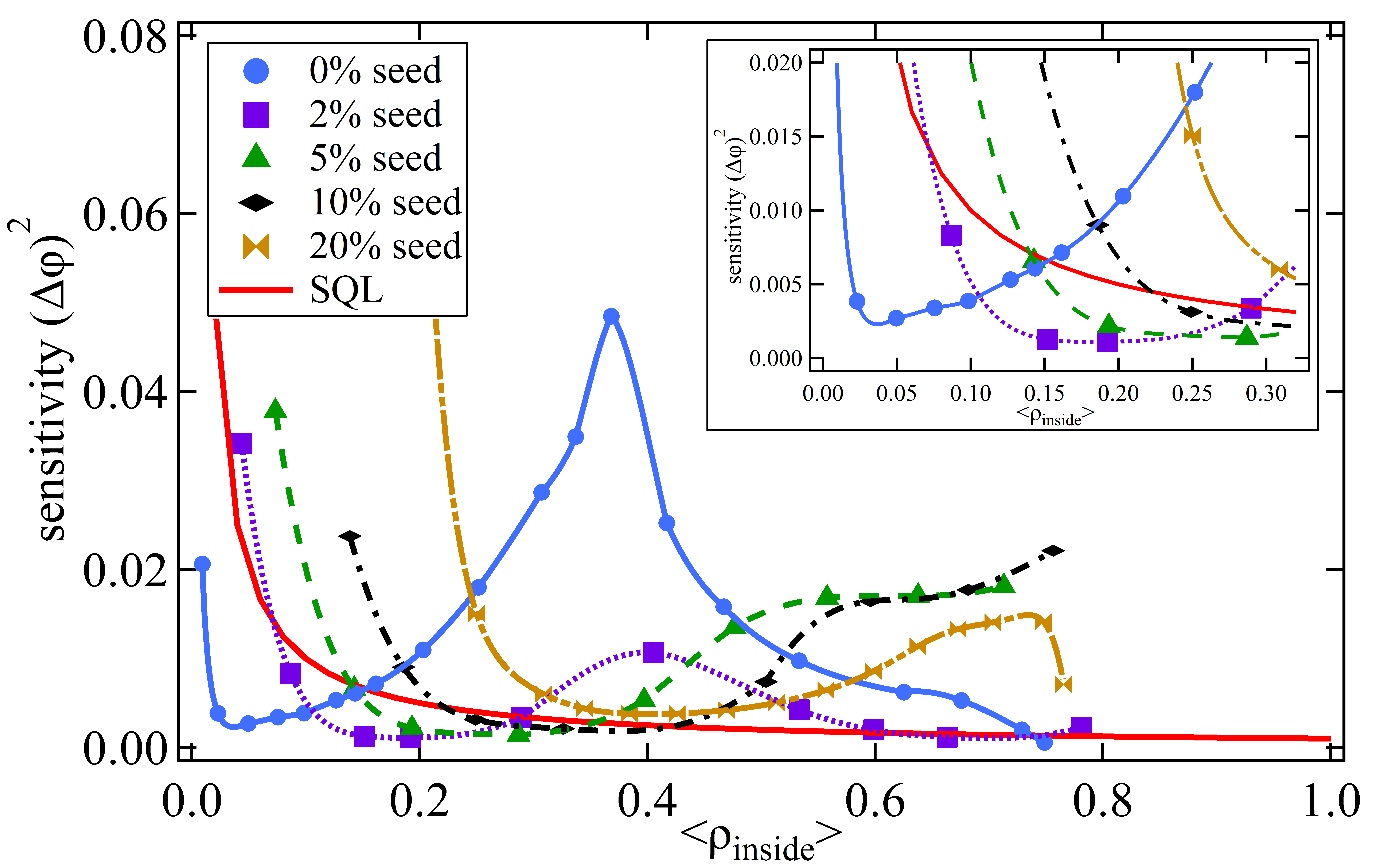}\label{fig:seeds_qeff2}
\caption{(Color online) Phase sensitivities for different initial seeds of 0\% (blue circles), 2\% (purple squares), 5\% (green triangles), 10\% (black diamonds), and 20\% (yellow crosses). Here, $N=1{,}000$ and $q/h=-2$~Hz. The inset shows a zoomed-in region where enhanced sensitivities are found. The SQL is shown as red solid line. Points below the red line correspond to quantum-enhanced sensitivities. The lines are intended as guide to the eye.}
\label{fig:seeds}
\end{figure}

\begin{figure}[h!]
\centering
\includegraphics[width=0.9\linewidth]{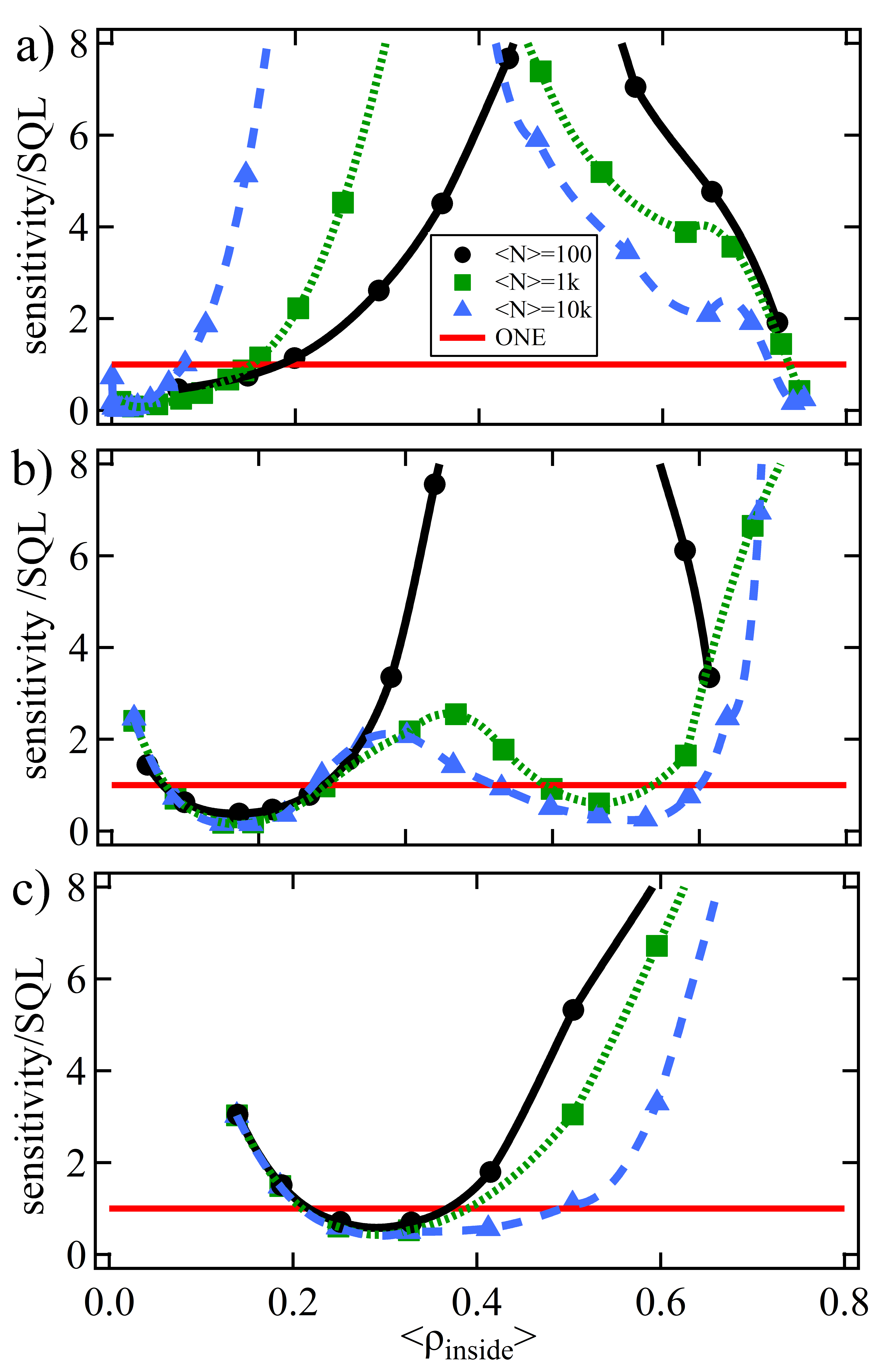}
\caption{(Color online) Phase sensitivities as a function of $\expect{\rho_{\textrm{inside}}}$ for different total atom numbers $N=100$ (black circles), $N=1{,}000$ (green squares), and $N=10{,}000$ (blue triangles), and initial seeds of (a) 0\%, (b) 2\%, and (c) 10\%. Here, ${q/h=-2}$~Hz. The red line depicts the ratio of sensitivity/SQL = 1. Points below the red line correspond to quantum-enhancement. Even for large atom numbers of $N=10{,}000$, quantum-enhancement is still predicted at longer evolution times. The lines are intended as guide to the eye.}
\label{fig:0-2-10seed}
\end{figure}

We now turn to compare the interferometry sensitivities for different initial seeds of 0\%, 2\%, 5\%, 10\%, and 20\% of a fixed total atom number $N=1{,}000$, shown in Fig.~\ref{fig:seeds}. We find sensitivities better than the SQL with up to 10\% initial seeds and ${\expect{\rho_{\textrm{inside}}}}$ up to 0.34. We obtain quantum-enhanced sensitivities for much larger numbers of atoms in the arms of the interferometer compared to the unseeded cases.

In Fig.~\ref{fig:0-2-10seed}, we investigate the effects of total number $N=100$, $N=1{,}000$ and $N=10{,}000$, on phase sensitivity with different initial seeds. We observe a strong dependence of sensitivity on $N$. Quantum-enhancement is present for all atom numbers that we studied. With larger seeds, the optimum sensitivity is obtained at larger values of $\expect{\rho_{\textrm{inside}}}$.
 
\section{Conclusion and Outlook}
In conclusion, we numerically studied spin-mixing interferometry in microwave-dressed F=1 Bose-Einstein condensates using realistic parameters that are accessible in experiments. We investigated the role of long evolution times and seeded initial states. By starting with coherent initial states with dual classical seeds from 0\% to 10\% in $m_F=\pm1$, combined with long evolution times $t\gg h/c$, larger total atom numbers become accessible to realize interferometers with quantum-enhanced sensitivities. These interferometers rely on highly non-sinusoidal interferometer fringes. We are using the simulation results presented here as guidance in our current experiments. We anticipate these results to be useful for future quantum technologies in matter-wave quantum optics, such as quantum-enhanced sensors based on spinor BECs.

\begin{acknowledgments}
This material is based upon work supported by the National Science Foundation under Grant No. 1846965. We acknowledge fruitful discussions with Dr.~Eite Tiesinga from the National Institute of Standards and Technology, Gaithersburg, MD, and the Joint Quantum Institute, University of Maryland, College Park, MD, who also shared with us some of his computer code. We acknowledge many enlightening discussions with Prof. Doerte Blume, Dr. Qingze Guan, and Dr. Jianwen Jie from the University of Oklahoma, Norman, OK. Some of the computing for this project was performed at the OU Supercomputing Center for Education \& Research (OSCER) at the University of Oklahoma (OU).
\end{acknowledgments}

\bibliography{SpinorInterferometry}

\end{document}